\documentclass[conference]{IEEEtran}
\IEEEoverridecommandlockouts
\usepackage{cite}
\usepackage{amsmath,amssymb,amsfonts}
\usepackage{algorithmic}
\usepackage{graphicx}
\usepackage{textcomp}
\usepackage{xcolor}
\usepackage{graphicx}
\usepackage{amsmath}
\usepackage{fancyhdr} % 引入fancyhdr宏包

\newcommand{\Rmnum}[1]{\expandafter\@slowromancap\romannumeral #1@}

\begin{document}
	
	\title{What Roles Can Spatial Modulation and Space Shift Keying Play in LEO Satellite-Assisted Communications?\\
	}
	
	\author{  
		\IEEEauthorblockN{Chaorong Zhang\IEEEauthorrefmark{2}\textsuperscript{1},
			Qingying Wu\IEEEauthorrefmark{2}\textsuperscript{2},  
			Yuyan Liu\IEEEauthorrefmark{2}\textsuperscript{3}, 
			Benjamin K. Ng\IEEEauthorrefmark{2}\textsuperscript{4*},  
			and Chan-Tong Lam\IEEEauthorrefmark{2}\textsuperscript{5}}  
		\IEEEauthorblockA{\textit{\IEEEauthorrefmark{2}Faculty of Applied Sciences, Macao Polytechnic University, Macao SAR, China}
		\\  
			\textsuperscript{1}p2314785@mpu.edu.mo,  
			\textsuperscript{2}qingying.wu@mpu.edu.mo,
			\textsuperscript{3}p2311866@mpu.edu.mo, 
			\textsuperscript{4}bng@mpu.edu.mo,  
			\textsuperscript{5}ctlam@mpu.edu.mo
			\\
			\textsuperscript{*}Corresponding author: Benjamin K. Ng}  
	}

	\maketitle
	
	\begin{abstract}
		In recent years, the rapid evolution of satellite communications play a pivotal role in addressing the ever-increasing demand for global connectivity, among which the Low Earth Orbit (LEO) satellites attract a great amount of attention due to their low latency and high data throughput capabilities.
		Based on this, we explore spatial modulation (SM) and space shift keying (SSK) designs as pivotal techniques to enhance spectral efficiency (SE) and bit-error rate (BER) performance in the LEO satellite-assisted multiple-input multiple-output (MIMO) systems. 
		The various performance analysis of these designs are presented in this paper, revealing insightful findings and conclusions through analytical methods and Monte Carlo simulations with perfect and imperfect channel state information (CSI) estimation. 
		The results provide a comprehensive analysis of the merits and trade-offs associated with the investigated schemes, particularly in terms of BER, computational complexity, and SE. 
		This analysis underscores the potential of both schemes as viable candidates for future 6G LEO satellite-assisted wireless communication systems.
	\end{abstract}
	
	
	\section{Introduction}
	The advent of beyond 5th generation (B5G) wireless communications and the subsequent progression towards 6th (6G) wireless systems usher in an era where wireless communication systems are expected to provide higher data rates, lower latency, and improved connectivity. 
	One of the pivotal technologies in achieving these advancements is the utilization of satellite in communication systems, particularly those facilitated by Low Earth Orbit (LEO) satellites [1-3]. 
	LEO satellite-assisted wireless systems attract significant attention due to their potential to offer global coverage, reduced propagation delays, and enhanced signal quality compared to traditional Geostationary Earth Orbit (GEO) systems [4-6].
	This also makes LEO satellites particularly appealing for enabling advanced wireless communication systems that serve a wide range of applications, including broadband internet access, disaster recovery, and Internet of Things (IoT) connectivity [7-8].
	Nevertheless, LEO satellite-assisted wireless communications still require improvements in throughput and robustness to fully meet the demands of future 6G communications.
	\par
	In recent years, there is a surge of interest in the application of advanced signal processing and communication techniques to enhance the data rate and throughput in traditional multiple-input multiple-output (MIMO) wireless communications. 
	Among these, the spatial modulation (SM) scheme [9] and its variants, e.g., the space shift keying (SSK) scheme [10], emerge as promising strategies to improve the spectral efficiency (SE) and energy efficiency of such systems. 
	The SM scheme is a novel multiple-antenna technique that exploits the spatial dimension by dynamically activating transmit antennas to convey additional information, along with traditional constellation symbols modulated by $M$-ary phase shift keying (PSK)/quadrature amplitude modulation (QAM), to the receiver, thus providing higher SE [11].
	Besides, as inter-antenna interference and inter-channel interference at the transmitter are common challenges in MIMO wireless systems, the SM scheme effectively addresses these issues by activating only one transmit antenna in each time slot [12].
	However, the complexity of signal processing at the receiver and signal detection becomes higher compared to that of traditional MIMO schemes.
	To mitigate the higher complexity in the SM scheme, the SSK scheme offers a simplified alternative by eliminating the need for constellation symbols processing at the transmitter and receiver, while still maintaining satisfactory bit-error rate (BER) performance [13].
	Similarly, while the SSK scheme shares some benefits of the SM scheme's benefits, such as reduced interference, it struggles to achieve the higher SE that SM offers with the same number of transmit antennas.
	Although some novel variants of the SM and SSK schemes are proposed later [14-18], they often involve trade-offs among interference, complexity, and error performance, requiring further investigation to fully understand their potential benefits in various application scenarios.
	Consequently, considering the aforementioned advantages, the SM and SSK schemes are designed into LEO satellite-assisted wireless systems to enhance spectral efficiency (SE) and ensure reliable BER performance by reducing interference. 
	\par
	The main contributions of our works are given as follows: 
	1) In order to further improve the SE of traditional LEO satellite-assisted MIMO wireless system, the LEO satellite-assisted SM (LEO-SM) and SSK (LEO-SSK) schemes are designed in this paper. 
	This paper is the first to explore and discuss the performance of these schemes under imperfect channel state information (CSI), as a notable contribution to the field of 6G LEO satellite-assisted wireless communications.
	2) The analytical performance of SE and detection complexity is briefly presented in this paper, where some interesting and insightful findings are also given. 
	3) Monte Carlo simulations are applied in our works, where the simulation results show the superiority in the BER performance and SE in the proposed schemes. Besides, by comparing the simulation results between the LEO-SM and LEO-SSK schemes, we also obtain some interesting insights, which further confirms the necessity of the trade-off selection between these two schemes.

	\begin{figure}
		\centering
		\includegraphics[width=8.8cm,height=6.8cm]{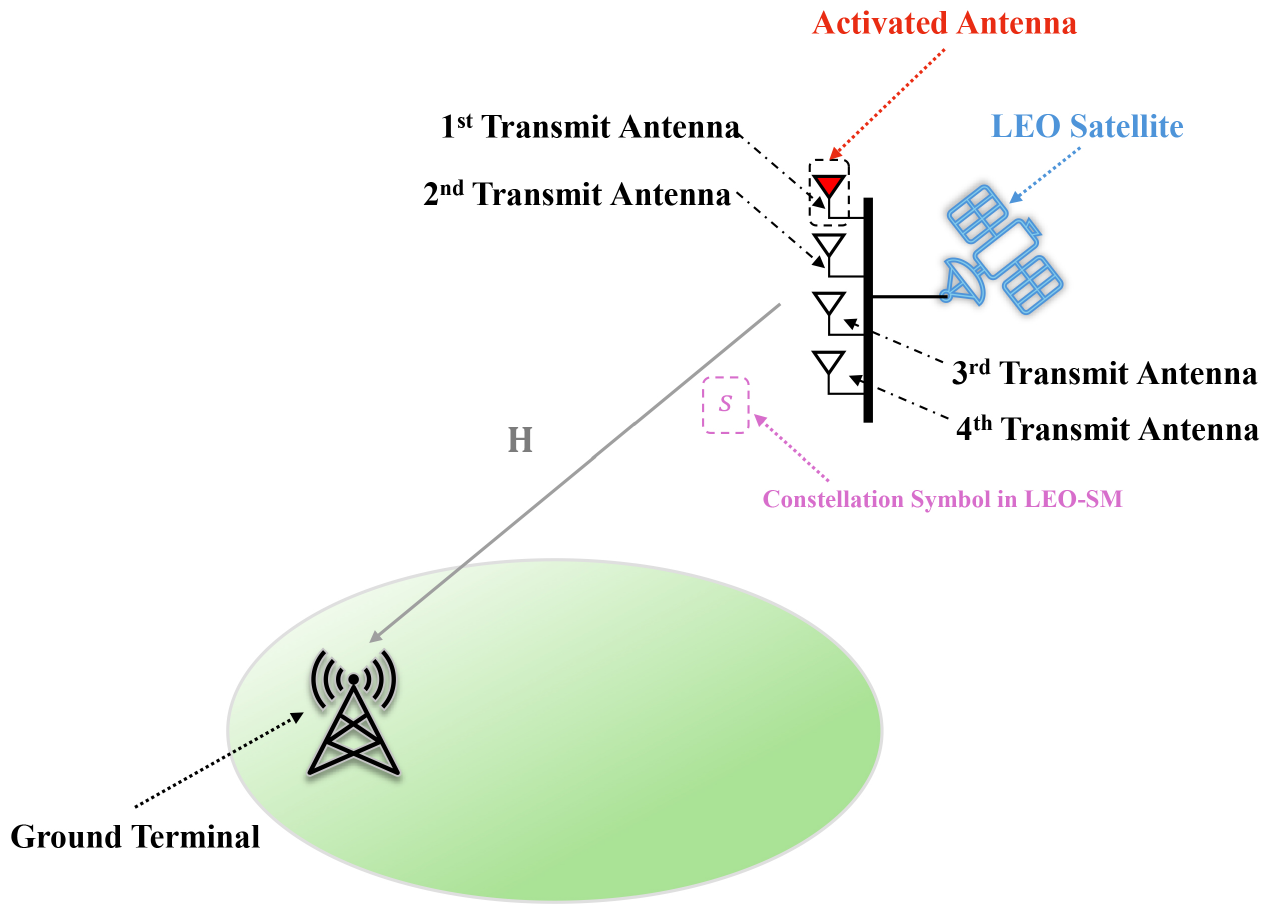}\\
		\caption{System model of the LEO-SM and LEO-SSK schemes.}
	\end{figure}
	
	\section{System Model}
	As shown in Fig. 1, we examine a wireless communication system facilitated by a LEO satellite, where a single satellite is equipped with $N_t$ transmit antennas, and a ground outdoor terminal is equipped with $N_r$ receive antennas. 
	$\footnote{While our study focuses on uncorrelated channel conditions for simplicity, the antenna correlation and correlated channels are thus not further extended in this paper, which will consider deeply analyzing in future works to provide a more comprehensive understanding.}$.
	In this configuration, the LEO satellite transmits signals to the ground terminal, which serves as the receiver [19].
	The ground terminal, located within the designated cell, can be viewed either as the user or the base station associated with the corresponding LEO satellite.
	To further enhance the SE of LEO satellite-assisted wireless networks, we design the SM scheme, where the bit sequence is mapped into $M$-ary PSK/QAM constellation symbols and the activated antenna is conveyed separately.
	Additionally, an SSK scheme is considered as an alternative, offering a trade-off that reduces complexity at both the transmitter and the receiver. 
	A detailed illustration of the processing involved in both schemes is provided in the following sub-subsection.
	
	\begin{figure*}
		\centering
		% Requires \usepackage{graphicx}\
		\includegraphics[width=16.5cm,height=10cm]{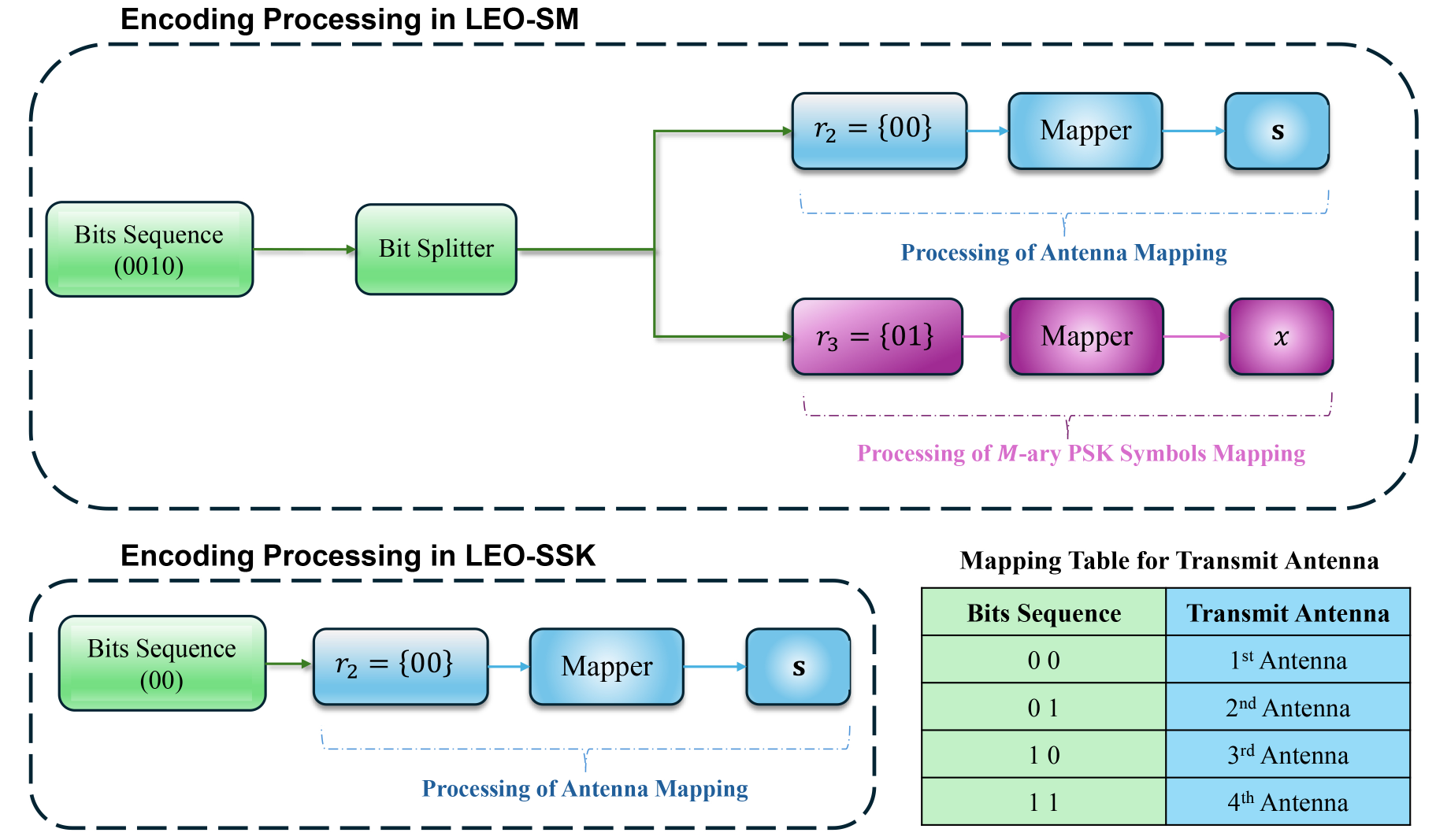}\\
		\caption{Encoding processing in the LEO-SM and LEO-SSK schemes.}
	\end{figure*}
	
	\subsection{Encoding Process}
 	In the traditional SM scheme, each transmit antenna represents a different bit sequence, and only one transmit antenna is selected to activate.
 	Through this activated transmit antenna, $M$-ary PSK/QAM symbol is transmitted to the receiver, wherein the receiver can obtain two sequences of transmitted bits separated from the antenna and constellation symbol.
 	For instance, according to Fig. 2 where the detail process and mapping table in SM and SSK encoding are given, transmitted bit sequence 0010 is divided into two sequences $r_1=\left\{ 00 \right\}$ and $r_2=\left\{ 10 \right\}$, which are mapped onto the quadrature PSK/QAM constellation symbol and first transmit antenna, respectively.
 	After this, $r_1$ and $r_2$ can be represented by a complex value of constellation symbol as $s$ and an antenna-selection vector as $\mathbf{v}=\left[ 1, 0, 0, 0 \right]$, where 0 represents the inactivated antennas, 1 represents the activated one, and the places of each element in $\mathbf{v}$ correspond to the positions and labels of each transmit antenna at transmitter in Fig. 1.
 	Multiplying $s$ and $\mathbf{v}$, the transmitted vector can be obtained, given as $\mathbf{x}_1=\left[ s, 0, 0, 0 \right]$.
	Without loss of generality, when $M$-ary constellation symbols are transmitted and $N_t$ transmit antennas are equipped, $\mathbf{v}$ and $\mathbf{x}_1$ can be denoted as $\mathbf{v}=\underset{N_t}{\underbrace{\left[ 0,1,..,0 \right] }}$ and $\mathbf{x}_1=\underset{N_t}{\underbrace{\left[ 0,s,..,0 \right] }}$.
	\par
	Similar to the encoding process of SM scheme, the antenna-selection vector in SSK scheme can be equally expressed as $\mathbf{v}=\left[ 0,1,..,0 \right] $.
	However, the constellation symbols are not involved in the SSK scheme, thus giving the transmitted vector as $\mathbf{x}_2=\left[ 0,1,..,0 \right]$.
	Fig. 2 also shows a specific example in encoding process of SSK scheme, where only $r_1=\left\{ 00 \right\}$ are conveyed and
	mapped into the antenna-selection vector as $\mathbf{v}=\left[ 1, 0, 0, 0 \right]$.
	It is worth noting that pure power is emitted from the activated transmit antenna, without any constellation modulation.
	
	\subsection{Channel Model}
	In this subsection, we detail the channel model for the forward terminal link in LEO satellite communication, divided into path loss and small-scale fading components [20-21].
	\par
	Firstly, the path loss, $L$, including large-scale effects, is expressed as:
	\begin{align}
		L = L_b + L_g + L_s,
	\end{align}
	where $L_b$, $L_g$, and $L_s$ represent the basic path loss, attenuation due to atmospheric gases, and attenuation from ionospheric or troposphere scintillation, respectively.
	The basic path loss in dB is modeled as:
	\begin{align}
		L_b = \mathrm{FSPL}(d, f_c) + \mathrm{SF} + \mathrm{CL}(\theta_E, f_c),
	\end{align}
	where $\mathrm{FSPL}$ is the free-space path loss, $\mathrm{SF}$ denotes shadow fading with Gaussian distributed as $\mathrm{SF} \sim \mathcal{N}(0, \sigma_{\mathrm{SF}}^2)$, and $\mathrm{CL}$ is the clutter loss. 
	The free-space path loss is calculated as:
	\begin{align}
		\mathrm{FSPL}(d, f_c) = 32.45 + 20 \log_{10}(f_c) + 20 \log_{10}(d),
	\end{align}
	with $d$ being the slant distance in meters, $f_c$ being the carrier frequency in GHz, and $\theta_E$ being the elevation angle from the user equipment (UE) to the satellite. The slant distance $d$ is determined by the satellite altitude $h_0$ and $\theta_E$ as:
	\begin{align}
		d = \sqrt{R_{E}^2 \sin^2(\theta_E) + h_{0}^2 + 2h_0R_{E}} - R_{E} \sin(\theta_E),
	\end{align}
	where $R_E$, the Earth's radius, is approximately 6,371 km.
	\par
	The attenuation due to atmospheric gases, dependent on frequency, elevation angle, altitude, and water vapor density, is given by:
	\begin{align}
		L_g = \frac{A_{zenith}(f_c)}{\sin(\theta_E)},
	\end{align}
	where $A_{zenith}$ represents the zenith attenuation for frequencies between 1 and 1000 GHz. 
	\par
	Scintillation loss $L_s$ arises from ionospheric or tropospheric effects, relevant for frequencies below 6 GHz (ionospheric) and above 6 GHz (tropospheric). 
	Rain and cloud attenuation are also considered for frequencies above 6 GHz, but are negligible under the clear-sky assumption used here.
	\par
	Secondly, the small-scale channel is modeled as a shadowed Rician fading channel with both line-of-sight (LoS) and non-LoS (NLoS) components. 
	We define $\mathbf{H}\in \mathbb{C} ^{N_r\times N_t}$ as the channel matrix between the LEO satellite and ground terminal.
	The shadowed Rician fading channel coefficient between $i$-th transmit antenna and $l$-th receive antenna is:
	\begin{align}
		h_{i,l} = \sqrt{\frac{K}{K+1}}|h_{i,l}^{LoS}| + \sqrt{\frac{1}{K+1}}|h_{i,l}^{NLoS}|,
	\end{align}
	where $K$ is the Rician factor and $\mathbf{H}=\left\{ h_{i,l} \right\} _{i=1,l=1}^{N_t,N_r}$.
	The LoS and NLoS components, $h_{i,l}^{LoS}$ and $h_{i,l}^{NLoS}$, follow:
	\begin{align}
		|h_{i,l}^{LoS}| \sim \mathrm{Nakagami}(m,\Omega), \quad \angle h_{i,l}^{LoS} \sim \mathrm{Unif}[0, 2\pi), \nonumber
		\\
		|h_{i,l}^{NLoS}| \sim \mathrm{Rayleigh}(\sigma_R), \quad \angle h_{i,l}^{NLoS} \sim \mathrm{Unif}[0, 2\pi),
	\end{align}
	where $m$ and $\Omega$ are the shape and spread parameters for the Nakagami distribution, and $\sigma_R$ relates to the average magnitude of the NLoS component.
	\par
	Thirdly, the Doppler shift $f_d$ [11-12] in the downlink can be given as:
	\begin{align}
		f_d=\frac{v}{c}\left( \frac{R_E}{R_E+h_0}\cos \alpha \right) f_c,
	\end{align}
	where $c$ is the speed of light and $v$ is the relative velocity between the satellite and ground station, $\alpha$ is the satellite elevation angle. 
	Meanwhile, by defining $D$ as distance between the center point of the satellite and ground terminal, we can obtain time delay $\tau$ as $\tau =\small{\frac{D}{c}}$.
	By considering the Doppler shift, the channel response from $i$-th transmit antenna to $n$-th reflection element and to user at $t$-th time slot can be respectively obtained as \footnote{If the one time slot is small or the mobility is slow enough, then $\tau_i$ can hold within a time slot $T_s$, i.e., the phase modification caused by frequency-dependent Doppler shift $e^{j2\pi \frac{f_d}{f_c} f t}$ within a frame can be ignored.}:
	\begin{align}
		h_{i,l}\left( t \right) =h_{i,l}e^{-j2\pi f_dt}e^{-j2\pi f_c\tau _t}\delta \left( \tau -\tau _t \right) ,
	\end{align}	
	where $\tau _t$ represents the delay at $t$-th time slot.
	Specifically, $\tau _t\simeq \eta \cdot t$ with $\eta =\frac{1}{3}\times 10^{-6}$ for vehicles moving at 360km/h.
	
	\subsection{Transmission and Detection}
	After obtaining the system configurations and channel models, the expression of received signals in the LEO-SM and LEO-SSK schemes can be given respectively as:
	\begin{align}
		\boldsymbol{y}_{LEO-SM}=\sqrt{L}\mathbf{Hv}s+\boldsymbol{n}
	\end{align}
	and
	\begin{align}
		\boldsymbol{y}_{LEO-SSK}=\sqrt{E_sL}\mathbf{Hv}+\boldsymbol{n},
	\end{align}
	where $\boldsymbol{n}\in \mathbb{C} ^{N_r\times 1}$ represents the white Gaussian noise vector with $\boldsymbol{n}=\left\{ n_l \right\} _{l=1}^{N_r}$, $n_l$ is white Gaussian noise with zero mean and variance $N_0$, and $E_s$ is the symbol's energy.
	\par
	Therefore, the instantaneous SNR at receiver in the LEO-SM and LEO-SSK schemes can be obtained as:
	\begin{align}
		\gamma _{l}^{LEO-SSK}=\sum_{i=1}^{N_t}{\sum_{l=1}^{N_r}{\sqrt{L}\left\| h_{i,l}(\mathbf{v}s-\hat{\mathbf{v}}\hat{s}) \right\| _2/N_0}}
	\end{align}
	and 
	\begin{align}
		\gamma _{l}^{LEO-SSK}=\sum_{i=1}^{N_t}{\sum_{l=1}^{N_r}{\sqrt{LE_s}\left\| h_{i,l}(\mathbf{v}-\hat{\mathbf{v}}) \right\| _2/N_0}}.
	\end{align}
	\par
	Besides, considering the accuracy in signals detection, the maximum likelihood (ML) detector is applied in both SM and SSK schemes, which can be respectively expressed as follows:
	\begin{align}
		\left\{ \hat{\mathbf{v}},\hat{s} \right\} =arg\underset{\mathbf{v},s}{\min}\left\| \boldsymbol{y}_{SM}-\sqrt{L}\mathbf{Hv}s \right\| _2,
	\end{align}
	and
	\begin{align}
		\left\{ \hat{\mathbf{v}} \right\} =arg\underset{\mathbf{v}}{\min}\left\| \boldsymbol{y}_{SSK}-\sqrt{E_sL}\mathbf{Hv} \right\| _2.
	\end{align}
	where $\mathbf{Hv}=\left\{ h_{i,l}^{\mathbf{v}} \right\} _{i=1,l=1}^{N_t,N_r}$, $\hat{\mathbf{v}}$ and $\hat{s}$ represent the estimated $\mathbf{v}$ and $s$.
	After the processing of ML detector, we can easily demodulate and decode the detected results, and obtain the transmitted bits sequences.
	Compared to Eq. (14) and (15), it is obvious that the complexity of ML detector in the LEO-SSK scheme is lower than the one in the LEO-SSK scheme, of which more details are illustrated in Sec. \uppercase\expandafter{\romannumeral3}.

	\subsection{Imperfect CSI Estimation} 
	In the practical satellite scenario, due to the complex wireless environment, the CSI estimation at receiver can hardly be perfect. 
	Considering the imperfect CSI estimation at receiver and referring to the imperfect CSI model from [22-23], the coefficient of $ h_{i,l}^{\mathbf{v}}$ can be expressed as the sum of $\dot{\mathrm{h}}_{i,l}^{\mathbf{v}}\sim \mathcal{C} \mathcal{N} (0,1-\delta _{e1}^{2})$ and $\ddot{\mathrm{h}}_{i,l}^{\mathbf{v}}\sim \mathcal{C} \mathcal{N} (0,\mathrm{ }\delta _{e1}^{2})$.
	Giving that $\delta_{e2}^2$ denotes the inaccuracy of estimation in CSI with $\ddot{\delta_{e2}^2}=1-\delta_{e2}^2$.
	Thus, we can obtain the receive signals between $t$-th transmit antenna and $l$-th receive antenna for the LEO-SM and LEO-SSK schemes with imperfect CSI estimation as
	\begin{align}
		\ddot{y}_{l}^{SM}=\sqrt{L}\left( \delta _{e2}^{2}\ddot{h}_{i,l}^{\mathbf{v}}+\ddot{\delta}_{e2}^{2}\dot{h}_{i,l}^{\mathbf{v}} \right) s+n_l
	\end{align}
	and
	\begin{align}
		\ddot{y}_{l}^{SSK}=\sqrt{E_sL}\left( \delta _{e2}^{2}\ddot{h}_{i,l}^{\mathbf{v}}+\ddot{\delta}_{e2}^{2}\dot{h}_{i,l}^{\mathbf{v}} \right) +n_l.
	\end{align}

	\section{Detection Complexity}
	The complexity of the ML detector in both the LEO-SM and LEO-SSK schemes is discussed in this section.
	We calculate the complexity based on Eq. (14) in the LEO-SM scheme and Eq. (15) in the LEO-SSK scheme, where the complex multiplications (CM) and complex additions (CA) are all required to count. 
	For the LEO-SM scheme, the operations required for CM and CA are detailed as follows:
	1) The multiplication $\mathbf{Hv}s$ involves $N_r\times N_t$ CMs and $(N_r\times (N_t- 1))$ CAs.
	2) The complex-scalar multiplication of the result $\sqrt{L}\mathbf{Hv}s$ by the complex symbol $s$ and real value $\sqrt{L}$ has $N_r$ CMs and 0 CAs
	3) The subtraction operation of $\boldsymbol{y}_{SM}-\sqrt{L}\mathbf{Hv}s$ involves $N_r$ CAs.
	4) Calculating the squared norm $\left\| \cdot \right\| _2$ requires $N_r$ CMs and $(N_r-1)$ CAs, where it needs to square the modulus and add up of each element, receptively.
	5) Totally, for each set $\left\{ \hat{\mathbf{v}},\hat{s} \right\}$, total complexity can be obtained as $N_r\left( N_t+2 \right) $ of CMs and $N_r\left( N_t+1 \right) -1$ of CAs.
	6) Finally, the total number of looping times by all set $\left\{ \hat{\mathbf{v}},\hat{s} \right\}$ is $N_t\times M$.
	\par
	Therefore, the total operations of CM and CA in ML detector for the LEO-SM scheme can be obtained as:
	\begin{align}
		C_{LEO-SM}=\left[ N_r\left( 2N_t+3 \right) -1 \right] N_tM.
	\end{align}
	Similarly, by following above steps, the total operations of CM and CA can be respectively obtained as $N_r\left( N_t+1 \right) $ and $N_r\left( N_t+1 \right) -1$ in $\left\| \boldsymbol{y}_{SSK}-\sqrt{E_sL}\mathbf{Hv} \right\| _2$ term of the LEO-SSK scheme, without the complex-scalar multiplication in term $\sqrt{E_sL}\mathbf{Hv}$. 
	Thus, the total operations of the ML detector applied in the LEO-SSK scheme can be obtained as:
	\begin{align}
		C_{LEO-SSK}=\left[ N_r(2N_t+2)-1 \right] N_t.
	\end{align}
	\par
	According to Eq. (18) and (19), we can easily find that the detection complexity increase with raise of $N_t$ in both proposed schemes and with $M$ arising in the LEO-SM scheme individually, which also indicates that the LEO-SM scheme has higher detection complexity than that of the LEO-SSK scheme when they share the same $N_t$ condition.
	Besides, we provide the running time by doing Monte Carlo simulations to show more comparisons and make more insightful conclusions in complexity analysis, which is presented and discussed in Sec. \uppercase\expandafter{\romannumeral5}.

	\section{Spectral Efficiency}
	Assuming $B_1$ bits are transmitted in the LEO-SM scheme, $B_1$ is divided into two parts, i.e., $b_1=\log _2N_t$ and $b_2=\log _2M$, which are mapped into the activated transmit antenna and onto the $M$-ary PSK/QAM constellation symbols, respectively.
	Specifically, $B_1$ can be expressed as follows:
	\begin{align}
		B_1 &= b_1+b_2 \nonumber
		\\
		&= \log _2N_t + \log _2M.
	\end{align}
	Here, for clarity, the SE performance is analyzed in bits per channel use (bpcu) across the LEO-SM and LEO-SSK schemes.
	Consequently, the SE of the LEO-SM scheme is given as $B_1$.
	Similar to $b_1$ of the LEO-SM scheme, the SE of the LEO-SSK scheme is given as $B_2 = \log _2N_t.$
	However, under the same conditions for $N_t$, the SE of the LEO-SSK scheme is lower than that of the LEO-SM scheme because the LEO-SM scheme allows for the additional transmission of constellation symbols.
	Besides, the SE of $B_1$ is reduced to $b_2$ when only constellation symbols are transmitted, as in the traditional LEO satellite-assisted $M$-ary PSK/QAM wireless transmission. 
	This reduction highlights the SE improvement offered by the LEO-SM scheme.

	 \begin{figure}
	 	\centering
	 	\includegraphics[width=8.6cm,height=6.6cm]{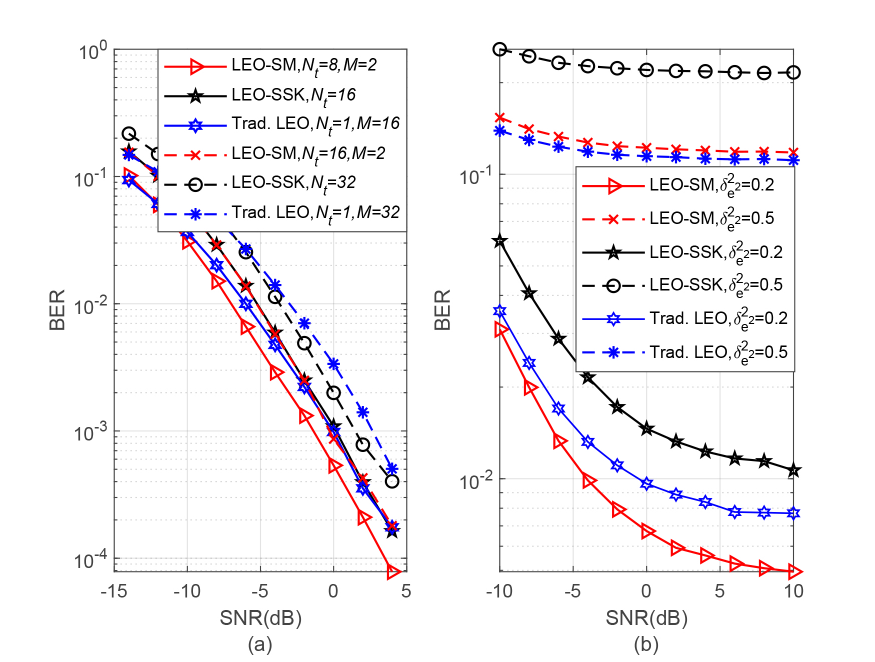}\\
	 	\caption{BER performance of the LEO-SM, LEO-SSK, and Trad. LEO schemes with $N_r=2$ and different $\delta_{e2}^2$.}
	 \end{figure}
	 
	 \begin{figure}
	 	\centering
	 	\includegraphics[width=8.6cm,height=6.6cm]{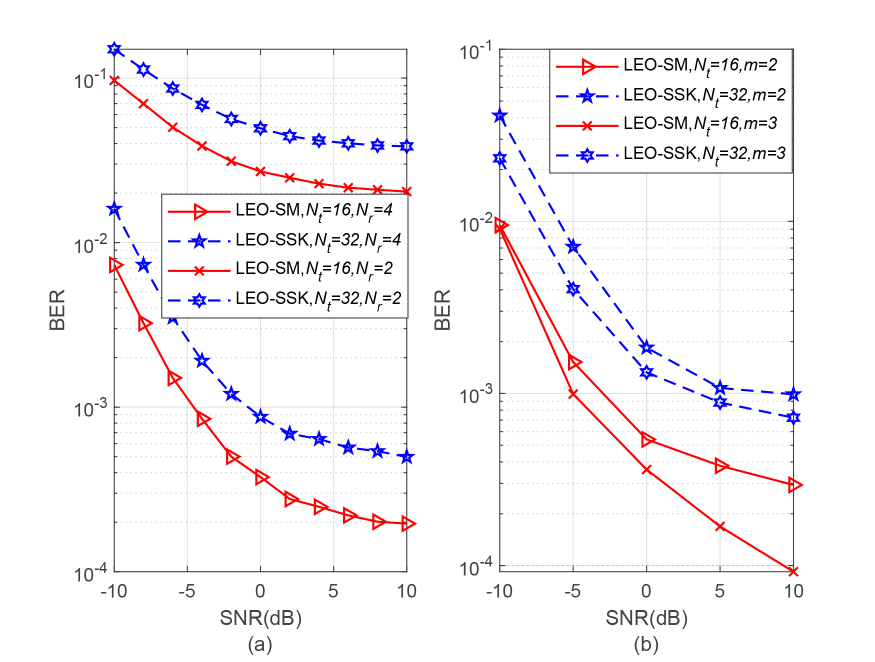}\\
	 	\caption{BER performance of the LEO-SM and LEO-SSK schemes with different $N_r$ and $m$.}
	 \end{figure}
	 
	 \begin{figure}
	 	\centering
	 	\includegraphics[width=8.6cm,height=6.6cm]{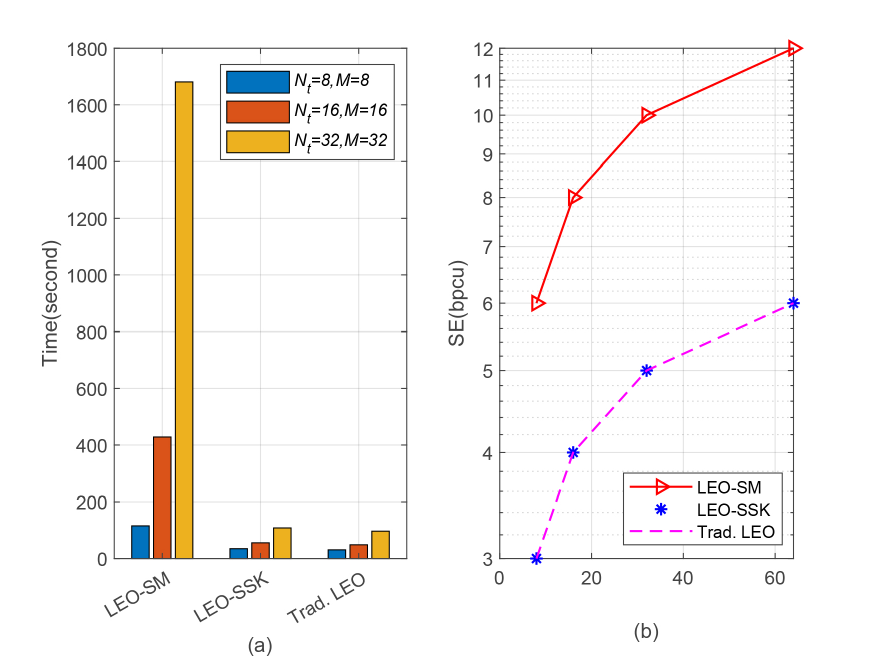}\\
	 	\caption{Complexity and SE of the LEO-SM, LEO-SSK, and Trad. LEO schemes.}
	 \end{figure}
	 
	\section{Theoretical and Simulation Results}
	The theoretical and simulations results of the proposed LEO-SM and LEO-SSK schemes are presented and discussed in this section, along with additional insights and conclusions.
	Monte Carlo simulations are considered in the simulations, with all experiments run for $1\times 10^6$ channel realizations.
	Similar to traditional schemes, $\frac{E_b}{N_0}$ is considered as the SNR, where $E_b=\left| x_k \right|^2=1$ represents the symbol's energy.
	The major simulation parameters are listed as follows [20-21]: 
	$f_c = 28 \, \text{GHz}$, 
	$h_0 = 780 \, \text{km}$, 
	$\theta_E = 60^\circ$, i.e., 
	$d = 884.85 \, \text{km}$, 
	$\sigma_{SF} = 1$, 
	$A_{zenith} = 0.22$, 
	$L_s = 0.13 \, \text{dB}$ (worst case), 
	$\sigma_R = 1$, 
	$\kappa = 1$, 
	$m = 0.8$, 
	and
	$\Omega = 1$. 
	For the parameters in Doppler shift, we let 
	$v= 100\, \text{km}$,
	$\alpha = 30^\circ$,
	and $D=700\, \text{km}$.
	We assume the LoS channel between LEO satellite and terminal, and then $\mathrm{CL}= 0$.
	Besides, the ML detector is employed in all simulation results of BER performance. 
	The traditional LEO satellite-assisted wireless communications scheme without any other spatial techniques is considered as the benchmark to make comparisons in this section, where the traditional one only transmit $M$-ary PSK/QAM constellation symbols and is called the Trad. LEO scheme for convenience.
	\par
	For fair comparison, the SE is equal to 4 bpcu for results in Fig. 3 (a) and 3 bpcu in Fig. 3 (b), with $N_r=2$. 
	Other different conditions of $N_t$, $M$, and $\delta_{e2}^2$ are given in Fig. 3, which are not elaborated in detail.
	The right figure illustrates the BER performance of the LEO-SM, LEO-SSK, and Trad. LEO schemes with $\delta_{e2}^2=0$.
	Fig. 3 (a) presents the BER performance through various non-zero values for $\delta_{e2}^2$, 
	As observed in Fig. 3 (a), the BER performance of the LEO-SM scheme outperforms that of the LEO-SSK and Trad. LEO schemes.
	This also indicates that the LEO-SM scheme has better BER performance than that of the LEO-SSK scheme under the high-speed mobile wireless scenario.
	Moreover, the traditional LEO scheme has the better BER performance than that of LEO-SSK scheme, which demonstrates that wireless signals only conveyed by antenna selection cannot provide satisfactory BER performance in LEO satellite-assisted communications.
	In Fig. 3 (b), the LEO-SM scheme shows superior BER performance compared to the LEO-SSK and Trad. LEO schemes under conditions of imperfect CSI estimation, specifically when $\delta_{e2}^2$ takes on values of 0.2 and 0.5, which indicates that the LEO-SM scheme has the better robustness with high $\delta_{e2}^2$.
	Besides, by comparing the results of the LEO-SM and LEO-SSK schemes across various values of $\delta_{e2}^2$, it is observed that the SNR at which the BER results shows the error floor increases as $\delta_{e2}^2$ diminishes.
	For a given $\delta_{e2}^2$, the LEO-SM scheme enters the error floor at a higher SNR value compared to the LEO-SSK scheme. 
	This observation further demonstrates the superior robustness of the LEO-SM scheme under imperfect CSI estimation conditions.
	\par
	Fig. 4 presents a comparative analysis of the LEO-SM and LEO-SSK schemes under the same values of $\delta_{e2}^2=0.2$ but with varying values of $N_r$ and $m$, as depicted in the left and right figures, respectively.
	Specifically, the LEO-SM scheme is characterized by $N_t=16$, $M=2$, and $N_r=2$, while the LEO-SSK scheme has $N_t=32$, $N_r=2$.
	As shown in Fig. 4, the BER performance of both schemes improves with an increase in $N_r$ and $m$,  and the SNR value at which the error floor occurs also increases with $N_r$ and $m$.
	\par
	Firstly, Fig. 5 has 4 sets of different conditions which are combined as $\mathbf{Q}=\left[ \left\{ 8,8 \right\} ,\left\{ 16,16 \right\} ,\left\{ 32,32 \right\} ,\left\{ 64,64 \right\} \right] $, with $\mathbf{Q}=\left\{ Q\left( q \right) \right\} _{q=1}^{4}$ and $Q\left( q \right) =\left\{ N_t,M \right\}$. 
	Since the complexity of the system is directly reflected by the running time of the simulations, running time of the LEO-SM, LEO-SSK, and Trad. LEO schemes are given in Fig. 5 (a), which has the conditions of $N_t$ and $M$ of first to third sets in $\mathbf{Q}$, giving as $Q\left( 1 \right) =\left\{ 8,8 \right\}$, $Q\left( 2 \right) =\left\{ 16,16 \right\}$, and $Q\left( 3 \right) =\left\{ 32,32 \right\}$.
	In the left figure, it is evident that the complexity of the schemes increases with the rise in $N_t$ and $M$. 
	Particularly in the LEO-SM scheme, due to the processing in constellation symbols and antenna selection meanwhile, the complexity is much larger than the one of others, especially when $N_t$ and $M$ are in large values.
	Conversely, the complexity of the LEO-SSK is slightly higher than that of the Trad. LEO schemes, which is because of the programming distinction and behaves with far fewer differences in practical applications.
	Also, the results of SE of these three scheme are presented in Fig. 5 (b), of which all sets in $\mathbf{Q}$ are considered.
	Apparently, the LEO-SM scheme demonstrates a significantly higher spectral efficiency (SE) compared to both the LEO-SSK and traditional LEO schemes, where the gap between the SE results of the LEO-SM scheme gets wider compared to those of the others as increase of $N_t$ and $M$, thereby highlighting the superior data rate and throughput of the LEO-SM scheme.
	Overall, as the trade-off selections designed in the LEO satellite-assisted wireless communication, the applications of LEO-SM and LEO-SSK schemes need to consider based on practical circumstances and requirements.
	Regardless, all results and analyses substantiate that both proposed schemes possess ample potential for application in future 6G wireless networks.

	\section{Conclusion}
	This study introduces and evaluates the performance of the LEO-SM and LEO-SSK schemes within the context of 6G wireless communications. 
	The applications of these advanced signal processing techniques are demonstrated to offer significant improvements in SE, robustness, and BER performance in the LEO satellite-assisted wireless systems, subject to both perfect and imperfect CSI estimation.
	As trade-off options, the LEO-SM and LEO-SSK schemes hold significant potential for enhancing future 6G wireless networks, particularly in scenarios requiring high data throughput and reliable connectivity, while the balance between complexity and performance requires careful consideration in practical deployment.

\end{document}